\documentclass[]{aastex6}

\usepackage{multirow}
\fullcollaborationName{The Friends of AASTeX Collaboration}

\begin{document}
\title{Likely GeV emission from an old Supernova Remnant: SNR G206.9+2.3}

\author{Yunchuan Xiang\altaffilmark{1} and Zejun jiang\altaffilmark{1}}
\altaffiltext{1}{Department of Astronomy, Yunnan University, and Key Laboratory of Astroparticle Physics of Yunnan Province, Kunming, 650091, China, xiang{\_}yunchuan@yeah.net, jiang{\_}zejun@yeah.net}


\begin{abstract}
A novel $\gamma$-ray supernova remnant (SNR) G206.9+2.3 is first reported in this study. We arrived at this conclusion after analyzing 12.4 years of observation data of the Fermi Large Area Telescope (Fermi-LAT).  The photon flux of the remnant was (1.19$\pm$0.59) 10$^{-9}$ cm$^{-2}$ s$^{-1}$, and 
its power-law spectral index was 2.22$\pm$0.19 in the 0.2-500 GeV energy band. 
Moreover, we found that the test statistic values of the global fit from the four different energy bands were greater than 9. 
We identified that this was a real $\gamma$-ray signal. 
Furthermore, we found that its GeV spatial location was in good agreement with that of its radio band. 
Its spectral energy distribution and light curve properties were similar to those of SNRs.  
We suggest that the novel $\gamma$-ray source is a likely counterpart to SNR G206.9+2.3. Consequently, we discuss its likely leptonic or hadronic origin.
\end{abstract}
\keywords{ supernova remnants - individual: (SNR G317.3-0.2) - radiation mechanisms: non-thermal}

\section{Introduction} \label{sec:intro}
SNR G206.9+2.3 (also known as PKS 0646+06) is an extended radio source classified as a supernova remnant (SNR) \citep{Clark1976}. \citet{Holden1968} stated that the source was detected at 178 MHz. 
Because this SNR was close to the Monoceros remnant \citep{Davies1978}, it was  regarded as an extended portion of Monoceros \citep{Caswell1970}. 
Subsequently, \citet{Day1972} confirmed that this was an independent object based on its 2650 MHz radio maps. 
\citet{Clark1976} used a revised flux density and distance relationship to derive a distance of 2.3 (kiloparsec) kpc.
 \citet{Davies1978} determined the fine filament structure of SNR G206.9+2.3 in the optical band. 

\citet{Nousek1981} provided the two upper limits of X-ray emission of SNR G206.9+2.3 based on HEAO-1 observations. 
In addition, they found that the X-ray intensity from the east and north portions        of SNR G206.9+2.3 and Monoceros Nebula all achieved the local maximum.
\citet{Graham1982} studied the extended features of SNR G206.9+2.3 at 2700 MHz using the Efferlsberg 100-m telescope and found that its 2700 MHz spectrum feature was highly consistent with those of the observed SNRs.  In addition, its contours at 2700 MHz were overlaid on the H$\rm \alpha$+[N$_{\rm \uppercase\expandafter{\romannumeral2}}$] map proposed by \citet{Davies1978}. Its distance range was estimated to be 3-5 kpc, and its shell-type morphology at 2700 MHz corresponds to the optical filamentary structure \citep{Graham1982}. 
\citet{Fesen1985} presented optical spectrophotometric data of SNR G206.9+2.3 
 with a 1.3 m telescope at the McGraw-Hill Observatory. They found that [O$_{\rm \uppercase\expandafter{\romannumeral1}}$] and [O$_{\rm \uppercase\expandafter{\romannumeral2}}$] line strengths for discriminating SNRs from [H$_{\rm \uppercase\expandafter{\romannumeral2}}$ region; moreover, they observed two thin filaments in its southern and northeastern regions. 
\citet{Leahy1986} found X-ray emissions of SNR G206.9+2.3 with Einstein. Using a Sedov model, they derived the age of the SNR to be approximately 60000 years and distance from the sun to be approximately 3-11 kpc. They estimated that this SNR was in an adiabatic blast phase based on a derived density of 0.04 cm$^{-3}$. 
\citet{Odegard1986} reported that the spectrum of SNR G206.9+2.3 has a turnover  between 20.6 MHz and 38 MHz; they believed that this turnover was likely to be caused by free-free absorption of cold and ionized gas. 
Using the OAN-SPM 2.1 m telescope, \citet{Ambrocio-Cruz2014} studied the kinematics of SNR G206.9+2.3 in the [SII]$\lambda$ 6717 and 6731 $\rm {\AA}$ emission lines. They estimated its distance to be 2.2 kpc from the sun, the explosion energy of this SNR was 1.7 $\times$ 10$^{49}$ ergs, and its age was 6.4 $\times$ 10$^{4}$ years in the radiative phase in the radiative phase.

Generally, SNR is widely regarded as the origin of high-energy cosmic rays \citep[e.g.,][]{Aharonian2004,Ackermann2013}. Astrophysical particles from the  cosmic rays of SNR can be accelerated to above 100 TeV energy bands by diffusive shock acceleration \citep[e.g.,][]{Aharonian2007,Aharonian2011}. In addition, the potential reacceleration processes within SNRs can accelerate these particles to the GeV/TeV energy band as well \citep[e.g.,][]{Caprioli2018,Cristofari2019}, which makes the energy band of SNR spectral energy distribution (SED) to 
be in the range of GeV/TeV observation band \citep[e.g.,][]{Zhang2007,Morlino2012,Tang2013}.
Thus far, 24 SNRs certified and 19 SNR candidates have been included in the  fourth Fermi catalog \citep{4FGL}. 
Previously, 
\citet{Acero2016b} did not find significant $\gamma$-ray radiation with Fermi Large Area Telescope (Fermi-LAT) for SNR G206.9+2.3. 
With the accumulation of photon events, in our preliminary exploration, a likely GeV $\gamma$-ray emission in the location of SNR G206.9+2.3 was found by checking its test statistic (TS) maps, which strongly encouraged us to further investigate its GeV features and likely origin of the SNR G206.9+2.3 region. This is of great significance in exploring the unknown origin of cosmic rays and the acceleration limit of particles within the SNR in the future.

The remainder of this paper is divided as follows: the data reduction is introduced in Section 2; the process of source detection is presented in Section 3; the likely origins of SNR G206.9+2.3 and its GeV features are discussed and summarized in Section 4.

\section{Data Preparation} \label{sec:data-reduction}
A binned likelihood tutorial\footnote{https://fermi.gsfc.nasa.gov/ssc/data/analysis/scitools/binned{\_}likelihood{\_}tutorial.html} was used in this analysis. 
Fermitool of version {\tt v11r5p3}\footnote{http://fermi.gsfc.nasa.gov/ssc/data/analysis/software/} was utilized in all subsequent analyses.
The photon events class with evclass = 128 and evtype = 3 and the instrumental response function (IRF) of ``P8R3{\_}SOURCE{\_}V3'' were selected to analyze the region of interest (ROI) of  $20^{\circ}\times 20^{\circ}$, which was centered at the location of (R.A., decl.= 102.17$^{\circ}$, 6.43$^{\circ}$; from SIMBAD\footnote{from http://simbad.u-strasbg.fr/simbad/}).
An energy range of 0.8-500 GeV was selected to maintain a small point spread function (PSF) and decrease the contribution from the galactic and extragalactic diffuse backgrounds.
The observation  period ranged from August 4, 2008 (mission elapsed time (MET) 239557427) to December 29, 2020 (MET 630970757). The maximum zenith angle of $90^{\circ}$ was selected to reduce the contribution from the Earth Limb. 
The script {\tt make4FGLxml.py}\footnote{https://fermi.gsfc.nasa.gov/ssc/data/analysis/user/} and the latest fourth Fermi catalog \citep[4FGL;][]{4FGL}, gll{\_}psc{\_}v27.fit\footnote{https://fermi.gsfc.nasa.gov/ssc/data/access/lat/10yr{\_}catalog/}, were used to generate a source model file, which included all sources within 30$^{\circ}$ around the SIMBAD location of SNR G206.9+2.3. 
 A point source with a power-law spectrum\footnote{https://fermi.gsfc.nasa.gov/ssc/data/analysis/scitools/xml{\_}model{\_}defs.html\#powerlaw} was subsequently added to the SIMBAD position of SNR G206.9+2.3 in the source model file to analyze its $\gamma$-ray features. We selected free spectral indexes and normalizations for all sources within the 5$^{\circ}$ range around the SIMBAD position of SNR G206.9+2.3. In addition,  
normalizations of two background models, including the isotropic extragalactic ({\tt iso{\_}P8R3{\_}SOURCE{\_}V3{\_}v1.txt}) and the galactic diffuse ({\tt iso{\_}P8R3{\_}SOURCE{\_}V3{\_}v1.txt})\footnote{http://fermi.gsfc.nasa.gov/ssc/data/access/lat/BackgroundModels.html.} diffuse backgrounds, were freed as well.

\subsection{\rm Source Detection} \label{sec:data-reduction}

\begin{figure*}[!h]
  \begin{minipage}[t]{0.495\linewidth}
  \centering
   \includegraphics[width=70mm]{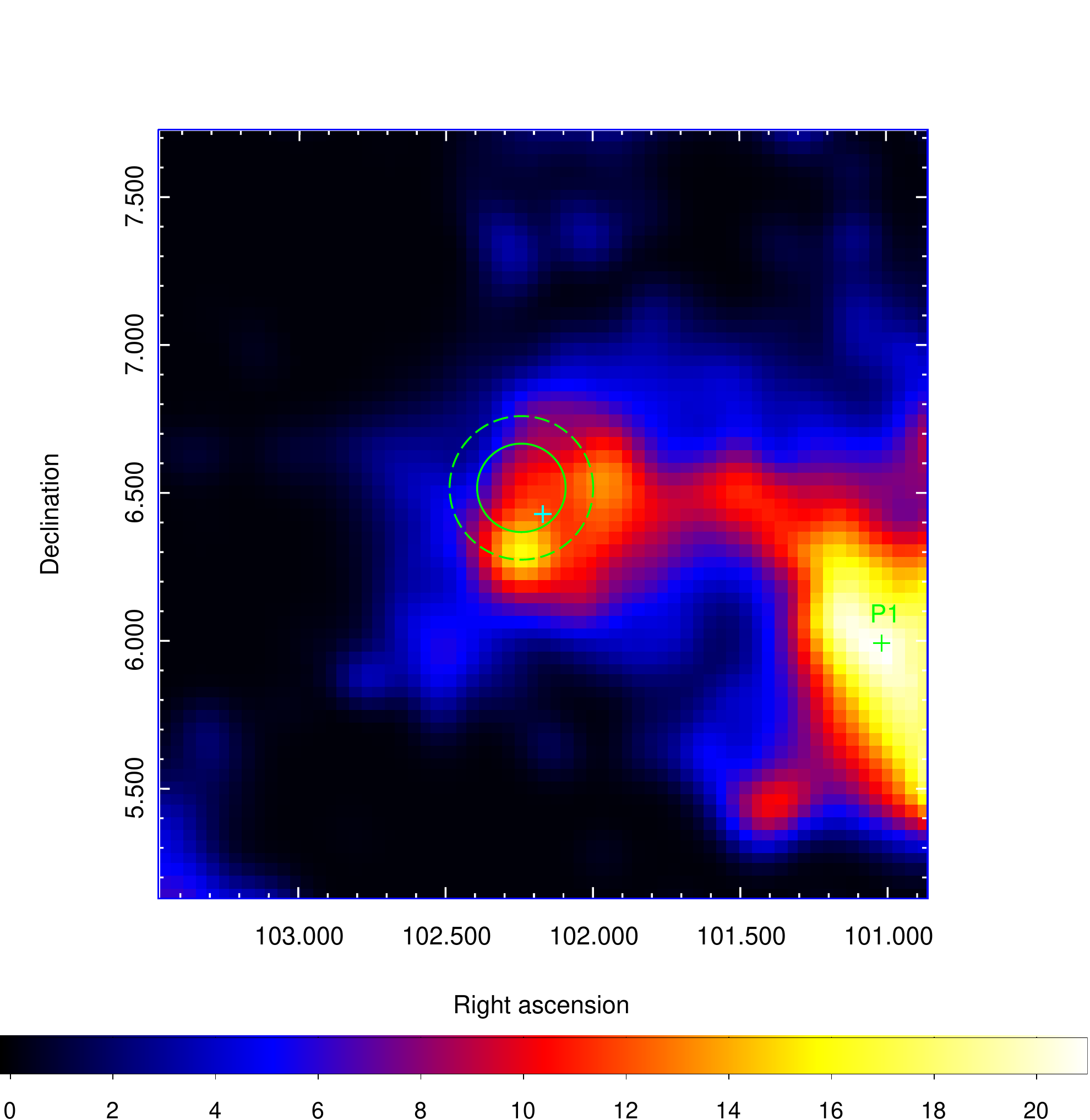}
   \centerline{(\emph{a})}
  \end{minipage}%
  \begin{minipage}[t]{0.495\textwidth}
 \centering
  \includegraphics[width=70mm]{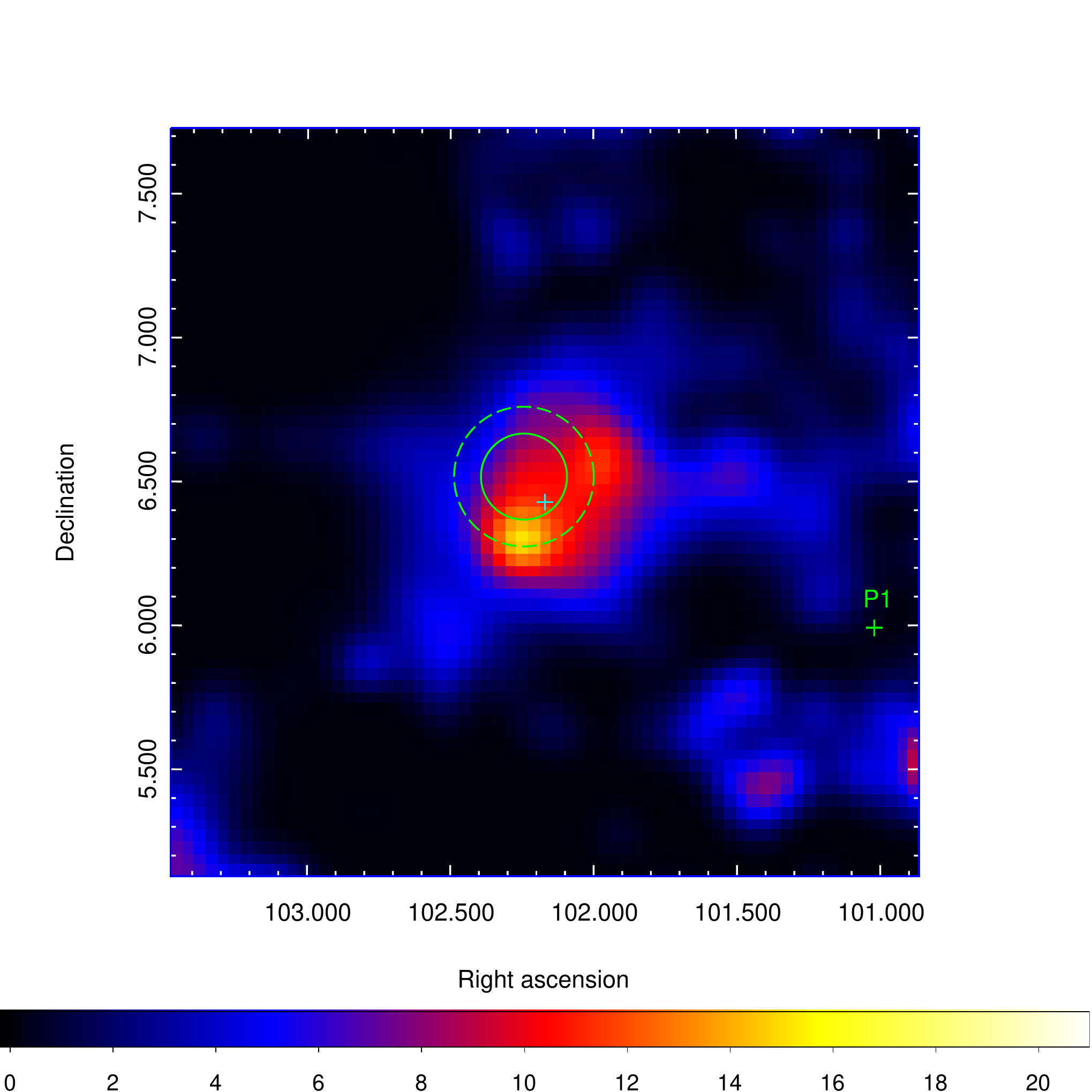}
    \centerline{(\emph{b})}
 \end{minipage}%
 \vfill
  \begin{minipage}[t]{1\linewidth}
  \centering
  \includegraphics[width=70mm]{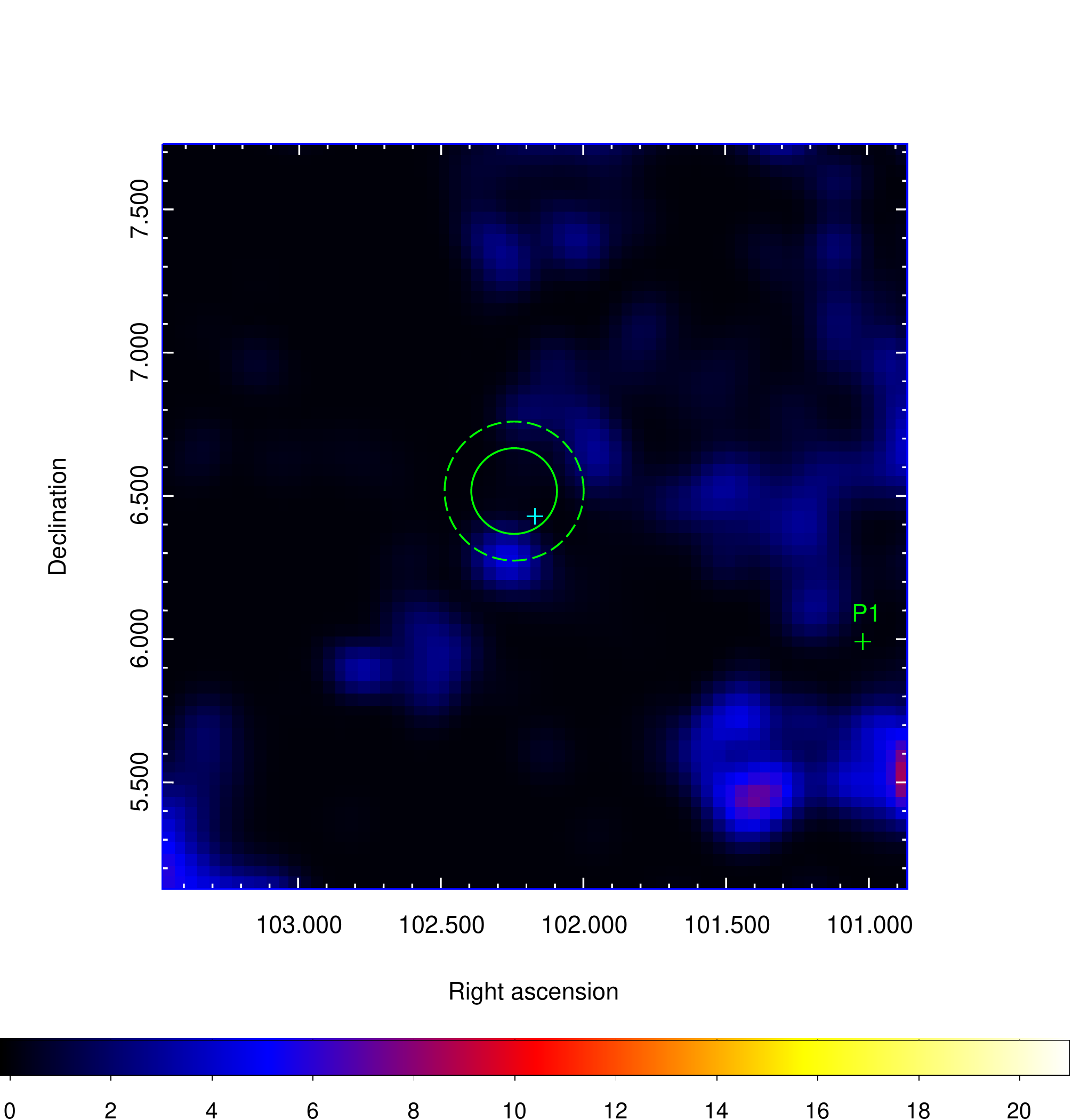}
    \centerline{(\emph{c})}   
 \end{minipage}%
\caption{
Size of the three TS maps is $2.6^{\circ} \times2.6^{\circ}$ with 0.04$^{\circ}$ pixel in the 0.8-500 GeV band. A Gaussian function with a kernel of $\sigma =0.3^{\circ}$ was used to smoothen them. The SIMBAD location of SNR G206.9+2.3 is  marked as a cyan cross. The 68\% and 95\% best-fit position uncertainties of SNR G206.9+2.3 are marked using solid and dashed green circles in each TS map, respectively.}
    \label{Fig1}
\end{figure*}

A complete TS map, including the $\gamma$-ray excess from the background of the SNR G206.9+2.3 region, was first generated using {\tt gttsmap}. As shown in panel (a) of Figure \ref{Fig1}, the SIMBAD location of SNR G206.9+2.3 showed  significant $\gamma$-ray radiation with a TS value of 12.27. 
To reduce the contribution from the surrounding $\gamma$-ray excess, we added a point source with a power-law spectrum to the location of P1 (R.A., decl.=101.02$^{\circ}$, 5.99$^{\circ}$) to exclude the $\gamma$-ray excess of the local maxima in the TS map for all subsequent analyses. Significant $\gamma$-ray radiation still existed and was more significant than in other locations in the 2.6$^{\circ}$ $\times$2.6$^{\circ}$ region, as shown in panel (b) of Figure \ref{Fig1}. 
Further, we excluded the emission of the SIMBAD position of SNR G206.9+2.3 to examine the probable $\gamma$-ray residual radiation near the location of SNR G206.9+2.3. However, we did not find any probable sources, as shown in panel (c) of Figure \ref{Fig1}. Its $\gamma$-ray radiation still existed and was significant with TS value of 10.87.

\begin{figure}[!h]
\centering
  \includegraphics[width=\textwidth, angle=0,width=110mm,height=100mm]{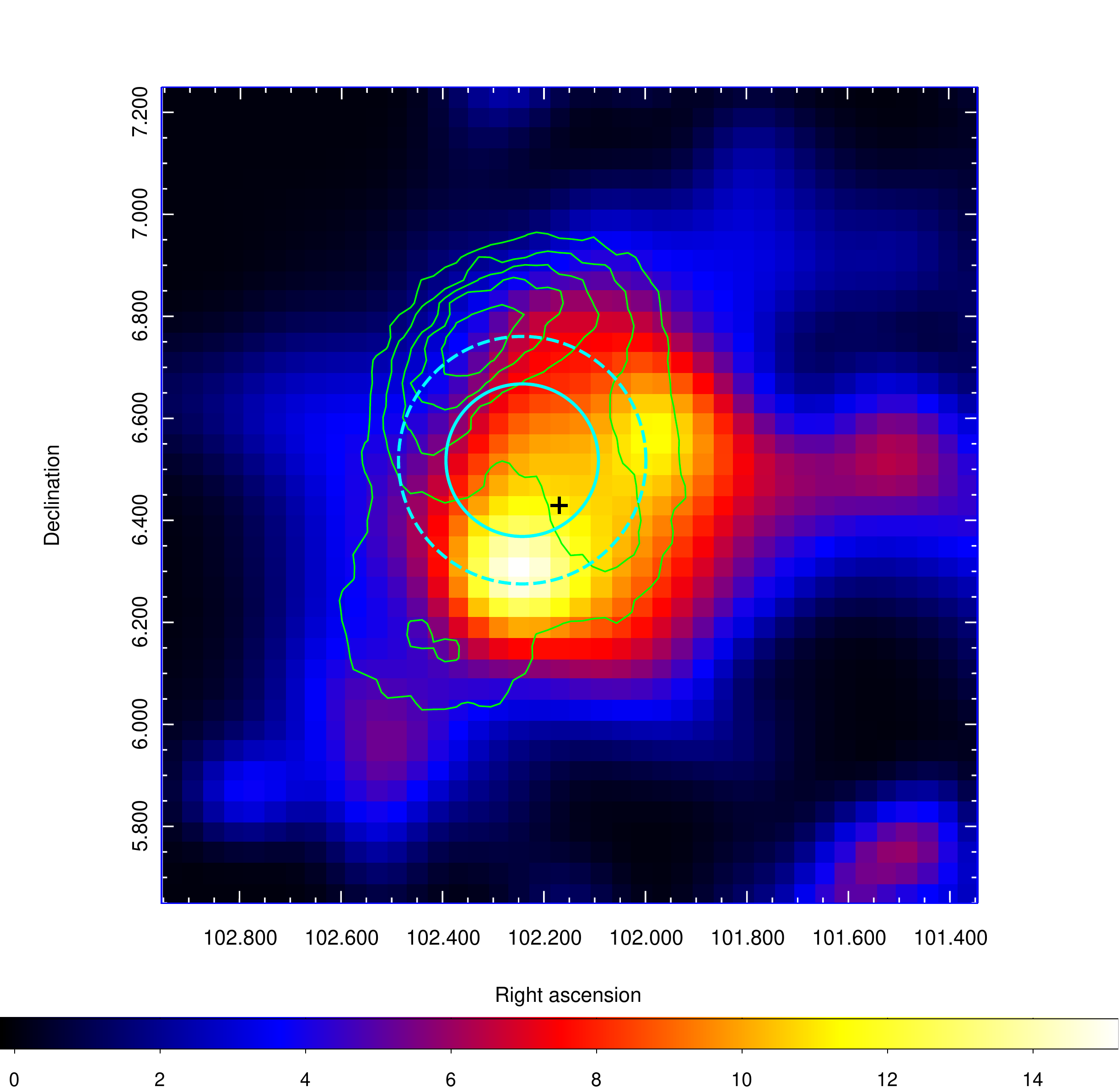} %
 \caption{Size of the TS map is $1.6^{\circ} \times 1.6^{\circ}$ with $0.04^{\circ}$ pixel in the 0.8-500 GeV energy band. Green contours are from the radio observation of the Effelsberg 100-m telescope \citep{Reich1997}. 
For other general descriptions, please refer to Figure \ref{Fig1}.
 }
 \label{Fig2}
\end{figure}

Furthermore, we calculated the best-fit position of the $\gamma$-ray radiation (R.A., decl. = 102.24$^{\circ}$, 6.52$^{\circ}$) with a 1$\sigma$ error circle of 0.14$^{\circ}$ using {\tt gtfindsrc}; we found that its SIMBAD position is within the 1$\sigma$ error circle of its best-fit location. Moreover, radio contours of SNR G206.9+2.3 from the Effelsberg 100-m telescope \citep{Reich1997} overlapped with the region within 2$\sigma$ error circle. Therefore, we suggest the discovered $\gamma$-ray source is likely to be a counterpart of SNR G206.9+2.3, as shown in Figure \ref{Fig2}.

Subsequently, the $\gamma$-ray spatial distribution of SNR G206.9+2.3 in the 0.8-500 GeV band was tested using the uniform disk and two-dimensional (2D) Gaussian models\footnote{https://fermi.gsfc.nasa.gov/ssc/data/analysis/scitools/xml{\_}model{\_}defs.html\#MapCubeFunction}.
Here, we used different radii and $\sigma$, ranging from 0.05$^{\circ}$ to 1$^{\circ}$, with a step of 0.05$^{\circ}$, to test results from these models. Calculating the values of TS$_{\rm ext}$, which is defined as 2log($L_{\rm ext}$/$L_{\rm ps}$), where $L_{\rm ps}$ and $L_{\rm ext}$ are the maximum likelihood values from a point source and an extended source, respectively. 
The maximum values of TS$_{\rm ext}$ from the uniform disk and 2D Gaussian models  were close to 4 and 3, respectively. 
Here, we present the best-fit results with the largest TS values from the two spatial models in Table \ref{table1}.
These results suggest a certain degree of performance for the spatial distribution of GeV $\gamma$-ray emission from the SNR G206.9+2.3 region.
Therefore, we chose the uniform disk model as the best-fit spatial model of the signal observed for subsequently spectral and timing analyses\footnote{Since the real spatial distribution of gamma rays is unknown for the SNR, and this object is relatively weak at present, we believe that such substitution is valuable for this study \citep[e.g.,][]{Feng2019}.}.

\begin{figure*}[!h]

      \begin{minipage}[t]{0.495\linewidth}
  \centering
  \includegraphics[width=66mm]{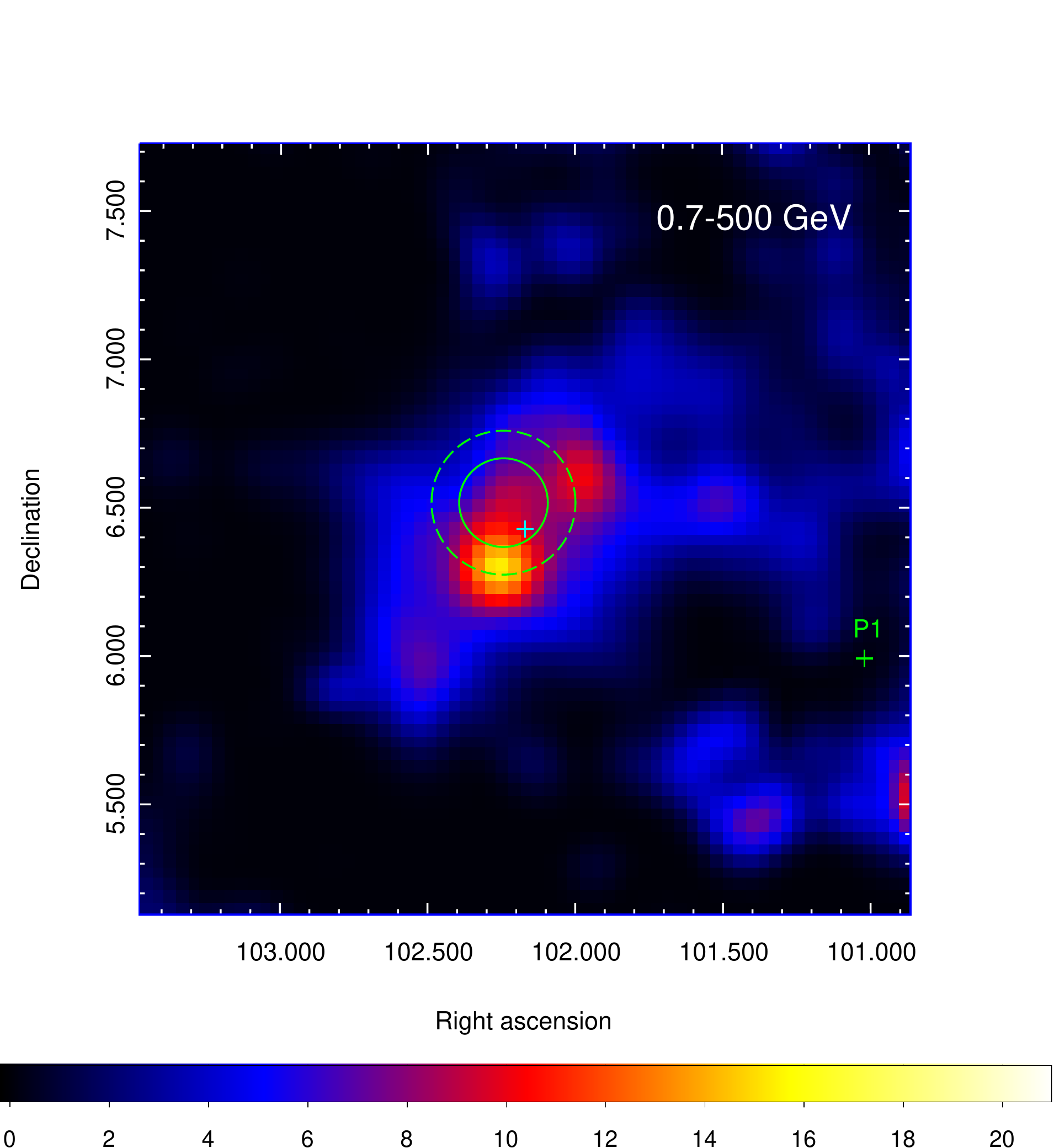}
       \centerline{(\emph{a})}
    \end{minipage}%
        \begin{minipage}[t]{0.495\linewidth}
     \centering
    \includegraphics[width=70mm]{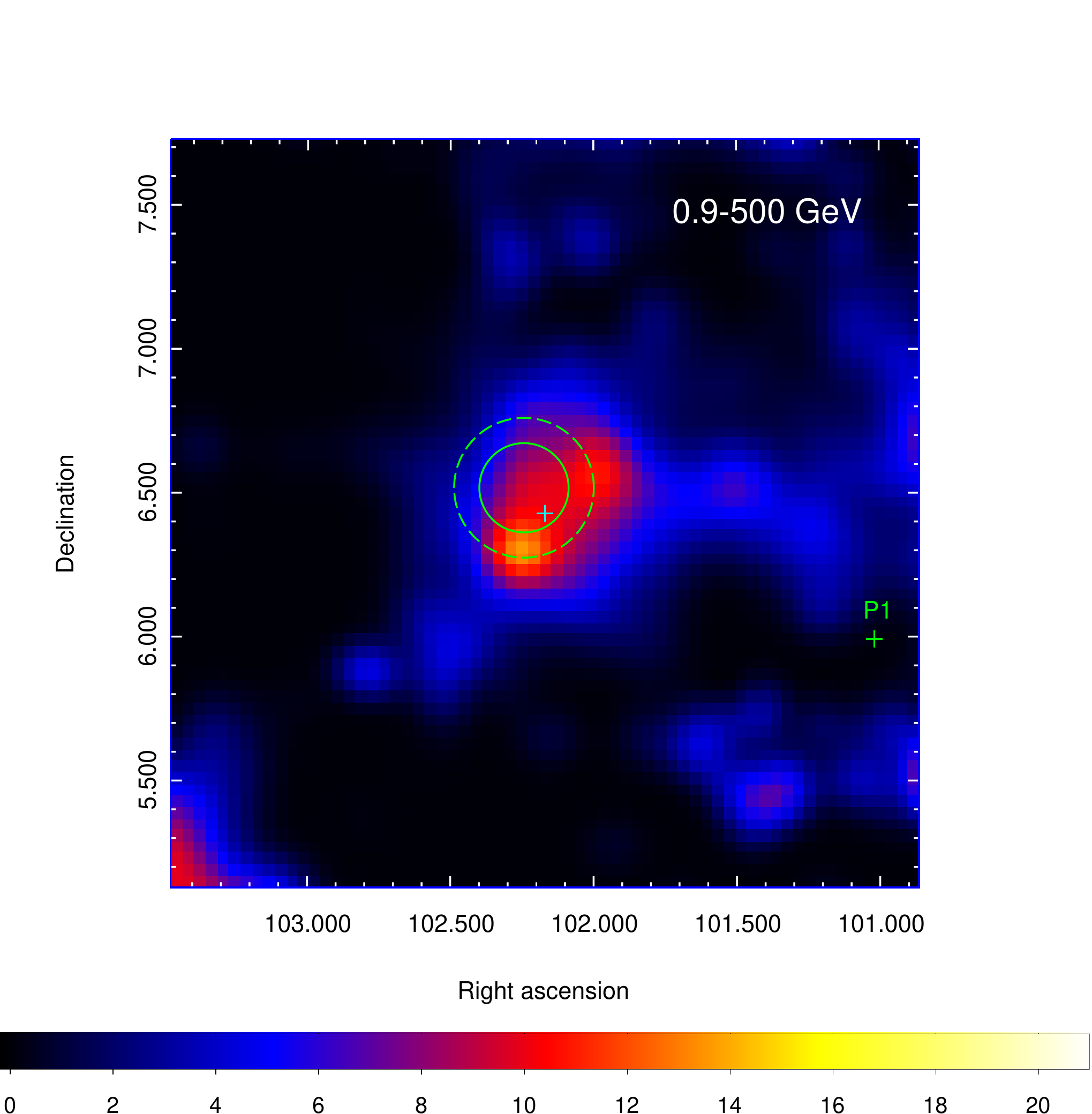}
     \centerline{(\emph{b})}
    \end{minipage}%
    
  \begin{minipage}[t]{1\linewidth}
  \centering
  \includegraphics[width=70mm]{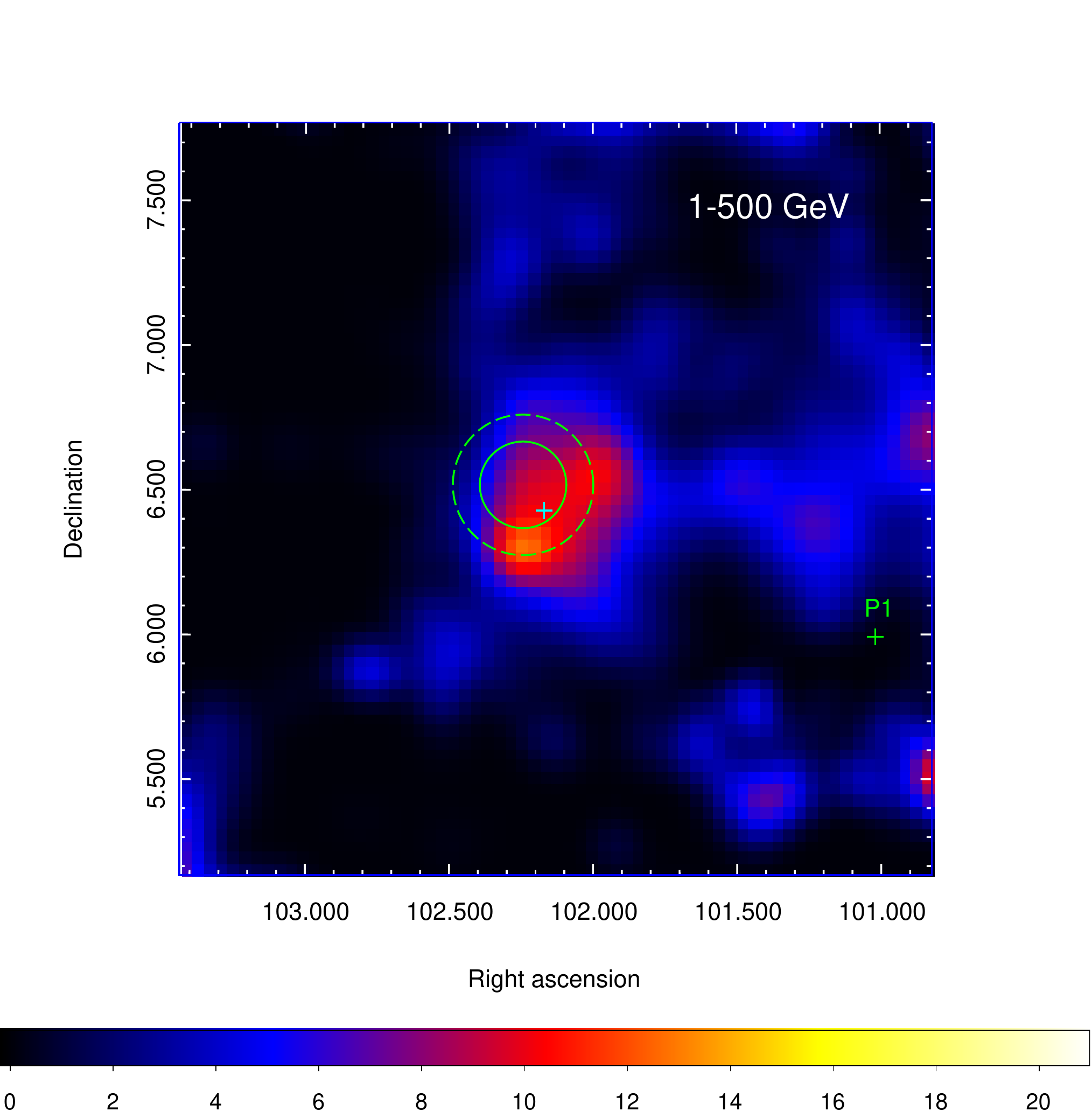}
       \centerline{(\emph{c})}
    \end{minipage}%

\caption{Three TS maps in different energy bands.
For other general descriptions, please refer to Figure \ref{Fig1}.
   }
    \label{Fig3}
\end{figure*}

To check the significance of this signal, we subsequently tested three other TS maps above 0.7 GeV, with an increment of 0.1 GeV. We found that the $\gamma$-ray residual radiation of SNR G206.9+2.3 is still more significant than other locations within the $2.6^{\circ}\times 2.6^{\circ}$ region, as shown in Figure \ref{Fig3}. 
Here, we present the correlative best-fit results of SNR G206.9+2.3 from the other three different energy bands with three kinds of spatial models in Table \ref{table2}. We found that their TS values were $>$9. Therefore, we proved that a new GeV source exists in the region. 
Although this source was weaker than most sources in 4FGL thus far, we cannot deny the authenticity of this GeV signal. For most weak GeV sources with TS$<$25, Fermi-LAT data point with TS value$>$4 can be regarded as credible data points  currently \citep[e.g.,][]{Xing2016,Xi2020b,Xiang2021a}. Based on SNR evolution, the age of SNR G206.9+2.3 is approximately 6.4 $\times$10$^{4}$ years \citep{Ambrocio-Cruz2014}, and it may be at a late evolution stage \citep{Guo2017}. 
 If SNR G206.9+2.3 enters the radiation cooling phase, as the velocity of the shock wave slows down continuously, the temperature of the material after the shock wave will rapidly cool, and the high-energy particles inside SNR will lose most energies through radiation. Meanwhile, it may be in a weak state above 1 GeV  \citep{Cox1972,Blondin1998,Brantseg2013}. 
For an old SNR G206.9+2.3, we thought that a low photon flux and TS value was  likely at present; thus, we suggested that the new GeV source is likely to be a counterpart of SNR G206.9+2.3.


%

\begin{table}[!h]
\begin{center}
\caption{The best-fit results for SNR G206.9+2.3 from different spatial models in 0.8-500 GeV band}
\begin{tabular}{lcccccccc}
    \hline\noalign{\smallskip}
    Spatial Model & Radius ($\sigma$) & Spectral Index  & Photon Flux & -log(Likelihood) & TS$_{\rm ext}$  \\
                  &     degree     &      &    $\rm 10^{-10}  ph$ $\rm cm^{-2} s^{-1}$  &  &     & \\       
  \hline\noalign{\smallskip}
   Point source    & ...             & 3.35$\pm$0.56 & 3.66$\pm$1.21 & 256046.27 & -  \\
   2D Gaussian        &  0.15$^{\circ}$  & 2.84$\pm$0.41 & 3.08$\pm$0.87 & 256044.60 & 3.34  \\
   uniform disk    &  0.25$^{\circ}$ & 2.81$\pm$0.39 & 4.48$\pm$1.30 & 256044.36 & 3.83     \\
  \noalign{\smallskip}\hline
\end{tabular}
    \label{table1} 
\end{center}
\end{table}

\begin{table}[!h]
 \scriptsize
\begin{center}
\caption[]{The best-fit parameters of SNR G206.9+2.3 for three different energy ranges}
\begin{tabular}{lclclclclclclc}
  \hline\noalign{\smallskip}
    \hline\noalign{\smallskip}
   Different Energy Range        & photon flux      & Spectral Model  &  $\Gamma$     &  TS value     \\
                   & $10^{-10}$ ph cm$^{-2}$s$^{-1}$ &        &               &               \\
  \hline\noalign{\smallskip}
   \multicolumn{5}{c}{point source model} \\
    \hline\noalign{\smallskip}
700 MeV - 500 GeV   & $3.74\pm1.41$  & PowerLaw         &  $2.96\pm0.42$        &  9.01    \\
900 MeV - 500 GeV   & $2.97\pm1.02$  & PowerLaw         & $3.49\pm0.71 $        &  10.00   \\
1000 MeV - 500 GeV  & $2.60\pm0.88$  & PowerLaw          &  $3.74\pm0.80$    & 10.32   \\
   \hline\noalign{\smallskip}
  \multicolumn{5}{c}{uniform disk model} \\ 
   \hline\noalign{\smallskip}
   700 MeV - 500 GeV   & $5.00\pm1.52$ & PowerLaw         &  $2.67\pm0.31$       &  14.29   \\
900 MeV - 500 GeV   & $3.61\pm1.11$  & PowerLaw         & $2.74\pm0.38 $        &  13.71   \\
1000 MeV - 500 GeV  & $3.19\pm0.98$  & PowerLaw          &  $2.90\pm0.46$    & 13.13   \\ 
    \hline\noalign{\smallskip}  
  \multicolumn{5}{c}{2D Gaussian model} \\
     \hline\noalign{\smallskip}
700 MeV - 500 GeV   & $5.20\pm 1.58$ & PowerLaw         &  $2.69\pm0.33$        &  13.94  \\
900 MeV - 500 GeV   & $3.74\pm1.14$ & PowerLaw  & $ 2.77\pm0.40$        &  13.21   \\
1000 MeV - 500 GeV  & $3.30\pm1.00$  & PowerLaw          &   $2.94\pm0.50$    & 12.67   \\ 
   \hline\noalign{\smallskip}
\end{tabular} 
 \label{table2}
\end{center}     
\end{table}

\subsection{Spectral Energy Distribution}

\begin{figure}[!h]
\centering
 \includegraphics[width=\textwidth, angle=0, scale=0.6]{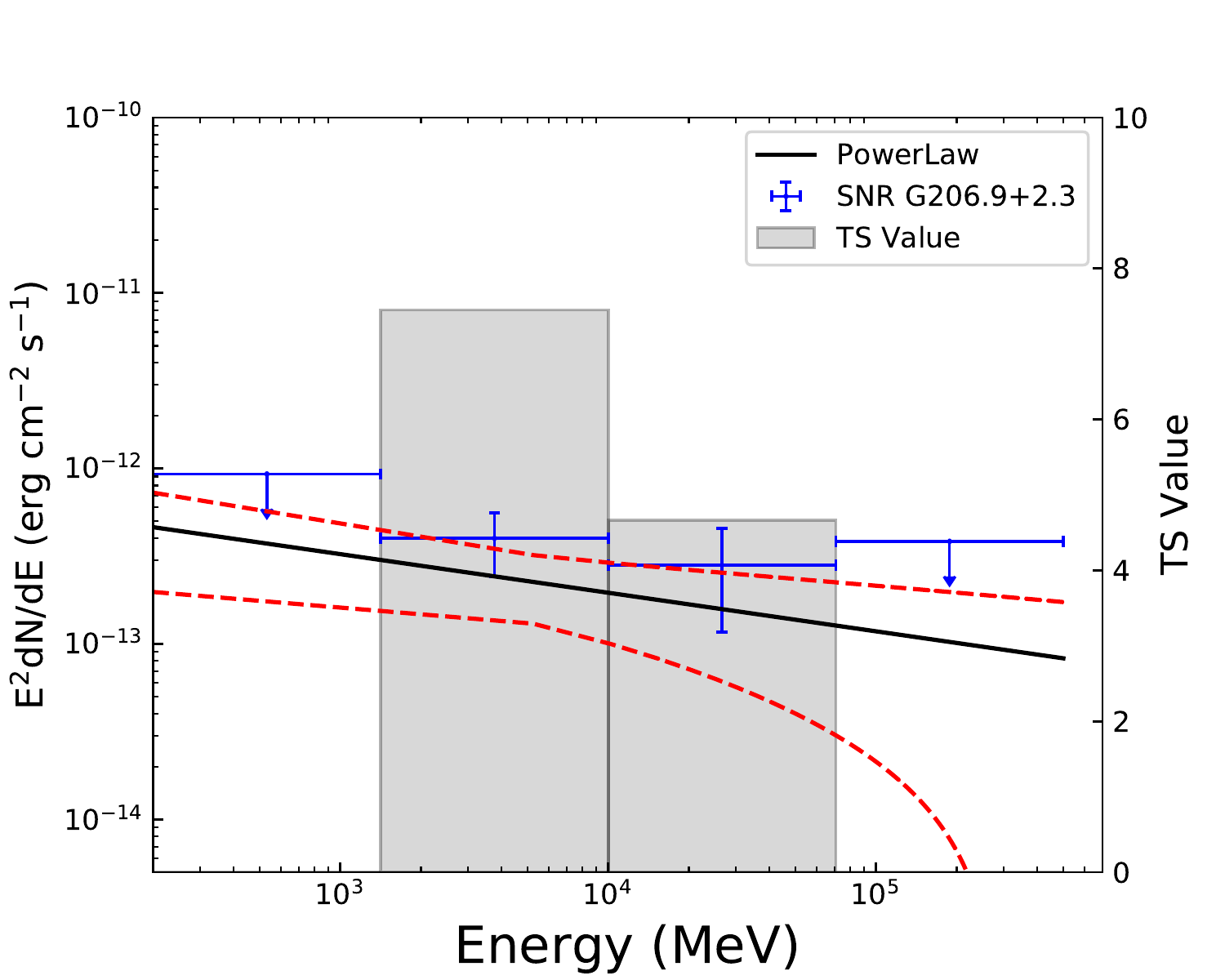}
 \caption{The blue data points are from this work. 
 TS values of above 4 are presented using a gray shaded region.
 The black solid line is the best-fit result of a power-law model, and the two red dashed lines are its 1$\sigma$ statistical uncertainties. 
 The 95\% upper limits are provided for data points with TS value $<$ 4.
 }
 \label{Fig4}
\end{figure}

In this analysis, we generated the spectral energy distribution (SED) with four equally logarithmic energy bins for SNR G206.9+2.3 in the 0.2-500 GeV bands. 
We found that the photon flux of the global fit is (1.19$\pm$0.59)$\times$ 10$^{-9}$  ph cm$^{-2}$ s$^{-1}$, and its spectral index is $2.22\pm 0.19$.  
Each energy bin of its SED was fitted separately using the binned-likelihood method. 
For the energy bin with a TS value $>$ 4, using the bracketing Aeff method\footnote{https://fermi.gsfc.nasa.gov/ssc/data/analysis/scitools/Aeff{\_}Systematics.html}, we calculated the systematic uncertainties from the effective area.  For the energy bin with a TS value $<$ 4, we assigned an upper limit with a 95\% confidence level, as shown in Figure \ref{Fig4}. The best-fit results for  each bin are presented in Table \ref{table3}.

\begin{table}[!h]
\begin{center}
\caption{The best-fit results of each energy bin of SNR G206.9+2.3}
\begin{tabular}{lcclclclc}
    \hline\noalign{\smallskip} 
    E  & Band       & $E^{2}dN(E)/dE$ & TS  \\
 (GeV) &  (GeV)     & ($10^{-13}$erg cm$^{2}s^{-1}$ ) &        \\            
  \hline\noalign{\smallskip}
    0.53    & 0.2-1.41    &  9.29 &   0.0  \\
    3.76    & 1.41-10.00    &  4.00$\pm$1.56$^{+0.19}_{-0.21}$  &  7.45      \\
    26.59  & 10.00-70.71    &  2.81$\pm$1.63$^{+0.62}_{-0.19}$  &  4.66   \\
    188.03 & 70.71-500   &  3.84  &  0.0  \\  
  \noalign{\smallskip}\hline   
\end{tabular}
    \label{table3}    
\end{center}
\textbf{Note}:
Two uncertainties from the second and the third energy fluxes are statistical and systematic ones, respectively. 
Other energy fluxes were the upper limits, with a 95\% confidence level.
\end{table}

\subsection{Variability Analysis}

\begin{figure}[!h]
\centering
 \includegraphics[width=\textwidth, angle=0,width=150mm,height=80mm]{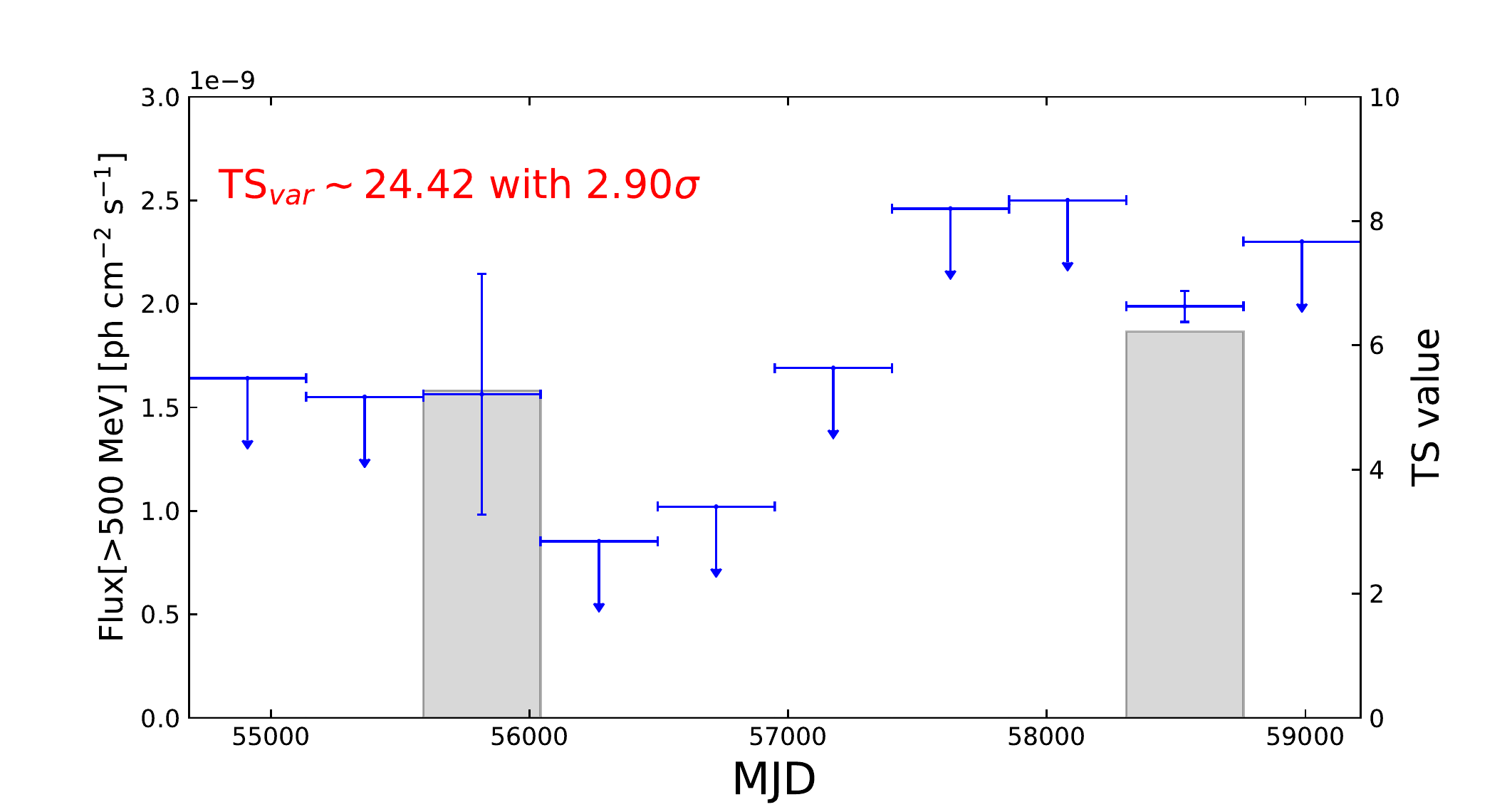} %
 \caption{Panel comprises 12.4 years of the LC with 10 time bins for SNR G272.2-3.2 in the 0.5-500 GeV band. 
The upper limits of the 95\% confidence level are calculated for the time bins  of TS values $<$ 4. The TS values are represented by gray shaded areas for the time bins with TS values $>$ 4.
}
 \label{Fig5}
\end{figure}

To decrease the pollution from the galactic diffuse background in the lower energy band, we chose an energy range of 0.5-500 GeV to generate a light curve (LC) with 10 time bins from the region of SNR G206.9+2.3, as shown in Figure \ref{Fig5}.  Subsequently, we calculated the value of $\rm TS_{var}$, as the variability index defined by \citet{Nolan2012}.  We found $\rm TS_{var} $ to be approximately equal to 24.42,   but greater than 21.67, which suggested that the LC of this SNR exhibits weak variability with a variability significance level of 2.90$\sigma$.

\section{discussion and conclusion}

\begin{table*}[!h]
\caption{The fit Parameters of Leptonic and Hadronic Models}
\begin{center}
\begin{tabular}{lcccccc}
    \hline\noalign{\smallskip}
  Model Name       &  $n_{\rm gas}$    & $\alpha$               & $E_{\rm cutoff}$       & $W_{\rm e}$ (or $W_{\rm p}$)    \\
                         &  (cm$^{-3}$) &                        & (TeV)                    & (erg)   &  \\
  \hline\noalign{\smallskip}
  Leptonic model    & --- & 1.50  & 0.76 &  1.04$\times 10^{47}$    \\
      \noalign{\smallskip}\hline 
  Hadronic model    & 0.1 & 2.30  & 11.82 &  1.13 $\times 10^{50}$  \\
   \hline\noalign{\smallskip}

\end{tabular}
\end{center}
\label{Table4}
Note: Here, for the distance and gas density of SNR G206.9+2.3, we assumed 2.2 kpc and 0.1 cm$^{-3}$ from \citet{Ambrocio-Cruz2014}, respectively. The energy band of the particles $>$ 1 GeV was selected to calculate the values of $W_{\rm e}$ and $W_{\rm p}$.
\end{table*}

In the beginning, we found that this $\gamma$-ray emission retained TS values$>$9 in different energy bands after reducing $\gamma$-ray contamination from surrounding significant residual radiations, so we suggested it was a real GeV $\gamma$-ray source. Then we found that the new $\gamma$-ray source was from the location of SNR G206.9+2.3; 
the GeV region of SNR G206.9+2.3 coincides with the radio region from the Effelsberg 100-m telescope. 
Using the simulation of the uniform disk model, we found 
its photon flux to be $\rm (3.68\pm 1.21)\times 10^{-10}$ $\rm cm^{-2} s^{-1}$ with a spectral index of $2.22\pm 0.19$ in the 0.2-500 GeV energy band, and the value of the spectral index is close to the average value of 2.15 of 43 SNRs and SNR candidates in 4FGL \citep{4FGL}.

In addition, we studied the spectral properties of other high-energy bands, and we found that the higher the energy band, the softer is the spectral index, as shown in Table \ref{table2}. 
For the elderly SNR G206.9+2.3, its particles inside may be in a late stage of evolution. At this stage, as the SNR shock wave continues to slow down, the high-energy particles inside will lose most of their energies through radiation, which makes its GeV spectrum likely present a soft property in the energy range of E$>$1 GeV \citep{Cox1972,Blondin1998,Brantseg2013}. 
 In addition, compared with other elderly SNRs, e.g., IC 443, W 44, W 51C, their ages are also above 10000 years, and the $\gamma$-ray spectrum also showed the soft property \citep{Guo2017}, and the spectral property of SNR G206.9+2.3 is similar to these of elderly SNRs currently observed.   
\citet{Tang2013} and \citet{Zeng2019} summarized the relationship between spectral cut-off energy ($E_{\rm cut-off}$) and SNR age. The $E_{\rm cut-off}$ will gradually decrease above 10$^{4}$ yr, which means that the high-energy photons inside SNR will gradually cool. Therefore, the cooling of the internal particles likely makes the current SNR G206.9+2.3 in a weak state, with a low photon flux and TS value.



Next, we studied its 12.4 years of LC; we found a weak variability with a variability significance level of 2.90$\sigma$. As shown in Figure \ref{Fig5}, we found that the TS values of the third and the ninth bins were higher than those of the  other bins. Subsequently, we immediately investigated other likely GeV candidates within the 2$\sigma$ error circle by SIMBAD, especially for common active galactic nuclei (AGN) or ANG candidates, but we did not find a likely candidate. Therefore, we thought the weak variability of this LC  likely originated from SNR G206.9+2.3 itself, similar to the newly discovered Supernova 2004dj \citep{Xi2020a,Ajello2020}. 

Leptonic and hadronic origins are widely used to explain the GeV $\gamma$-ray radiation of SNRs in the Milky Way \citep[e.g.,][]{Zeng2017,Zeng2019}. 
The former is generally considered to be caused by the inverse Compton scattering, and the latter is mainly owing to the decay of the neutral pion $\pi^{0}$ from the process of inelastic proton-proton collisions. 
For the hadronic origin, OH (1720 MHz) maser emission is considered to be convincing evidence for verifying the interaction of SNRs with OH molecular clouds. However, \citet{Frail1996} did not find significant OH maser emission from the SNR G206.9+2.3 region. 
In addition, \citet{Su2017b} made detailed observations of the CO molecular cloud around this SNR with the 13.7 m millimeter-wavelength telescope of the Purple Mountain Observatory. 
Although they did not find a wide CO molecular line from this region, this SNR was located in a molecular cavity at approximately 15 km s$^{-1}$; a certain number of CO molecular clouds were distributed around it, which implied a likely association between the surrounding CO molecular gas and the SNR.

\begin{figure}[!h]
\centering
 \includegraphics[width=\textwidth, angle=0,width=90mm]{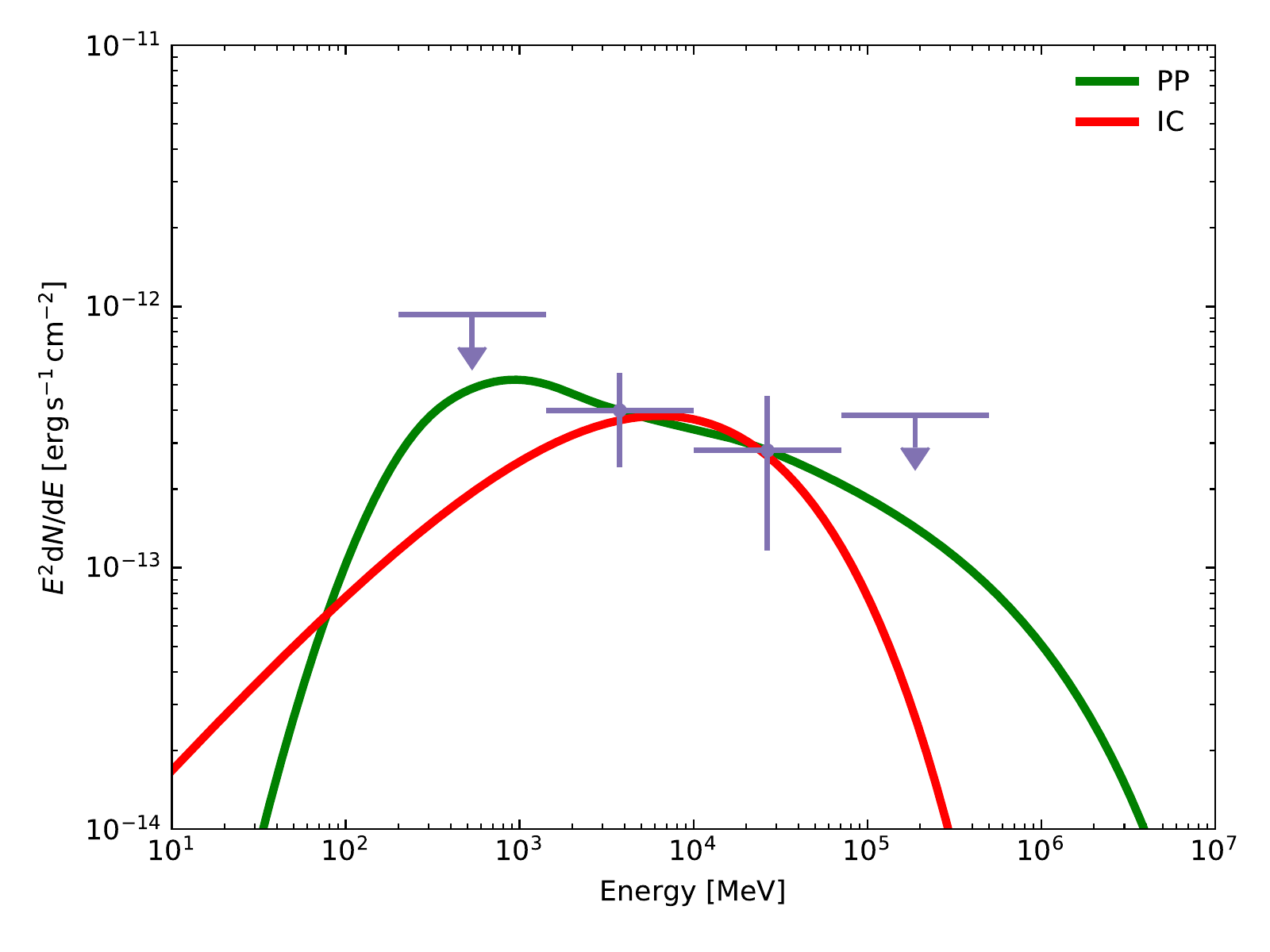} 
 \caption{Fit results of leptons and hadrons. The green solid line  represents the hadronic origin from the process of inelastic proton-proton collisions. The red solid line represents the leptonic origin from the inverse Compton scattering.
}
 \label{Fig6}
\end{figure}

Based on the aforementioned two kinds of origins, here we used a simple steady-state model provided by NAIMA \citep[][and references therein]{Zabalza2015} to explain its SED with the particle distribution of a power-law exponential cutoff, as was done by \citet{Xiang2021a}.  
We found that these two types of origins can explain the GeV SED of SNR G206.9+2.3, as shown in Figure \ref{Fig6}. The relevant fitting parameters are presented in Table \ref{Table4}. 
For the Leptonic fitting result, we found that an index of $\alpha$=1.5 was consistent with the average result of the four shell SNRs (including RX J1713.7-3946, RX J0852.0-4622, RCW 86, and HESS J1731-347; refer to \citet{Acero2015b}). Moreover, we found that particle electron energy budget $W_{\rm e}$($>$ 1 GeV) $\sim 10^{47}$ erg was in good agreement with that of \citet{Acero2015b}, which indicates that the observed radiation is probably of leptonic origin. 
For the hadronic fitting result, we found that its index of $\alpha$=2.3 was  consistent with the average result from SNRs with molecular cloud systems (SNR-MCs) (including IC 443, W 44, W 51C, W 49B, Puppis A; refer to \citet{Xiang2021a}). Furthermore, the proton energy budget ($W_{\rm p}$ $>$ 1 GeV) was in the  range of 10$^{49} \sim 10^{50}$ erg of SNR-MCs from  \citet{Xiang2021a}, which indicates that the hadronic origin of the radiation is also likely. 

Based on the aforementioned analyses, we propose that the new $\gamma$-ray source is likely to  be a counterpart of SNR G206.9+2.3. 
 More observational data from different wavebands are required to further reveal the origin of the $\gamma$-ray emission from the SNR G206.9+2.3 region in the future (e.g., continuous observations of Fermi-LAT).

\section{Acknowledgements}
We sincerely appreciate the support for this work from the National Key R\&D Program of China under Grant No.2018YFA0404204, the National Natural Science Foundation of China (NSFC U1931113, U1738211), the Foundations of Yunnan Province (2018IC059, 2018FY001(-003)), the Scientific research fund of Yunnan Education Department
(2020Y0039).


\begin{thebibliography}{}
 




\bibitem[Abdollahi et al.(2020)]{4FGL}Abdollahi, S., Acero, F., Ackermann, M., et al.   2020, ApJS, 247, 33

\bibitem[Acero et al.(2015b)]{Acero2015b}Acero, F., Lemoine-Goumard, M., Renaud, M., et al. 2015b, A\&A,580, A74

\bibitem[Acero et al.(2016b)]{Acero2016b}Acero, F., Ackermann, M., Ajello, M., et al.  2016b, ApJS, 224, 8

\bibitem[Ackermann et al.(2013)]{Ackermann2013}Ackermann, M., Ajello, M., Allafort, A., et al. 2013, Sci, 339, 807

\bibitem[Aharonian et al.(2007)]{Aharonian2007}Aharonian, F., Akhperjanian, A. G., Bazer-Bachi, A. R., et al. 2007, A\&A, 464, 235

\bibitem[Aharonian et al.(2004)]{Aharonian2004}Aharonian, F. A., Akhperjanian, A. G., Aye, K., et al. 2004, Natur, 432, 75

\bibitem[Aharonian et al.(2011)]{Aharonian2011}Aharonian, F., Akhperjanian, A. G., Bazer-Bachi, A. R., et al. 2011, A\&A, 531, C1

\bibitem[Ajello et al.(2020)]{Ajello2020}Ajello, M., Di Mauro, M., Paliya, V., \& Garrappa, S. 2020, ApJ, 894, 88

\bibitem[Ambrocio-Cruz et al.(2014)]{Ambrocio-Cruz2014}Ambrocio-Cruz, P., Rosado, M., Le Coarer, E., Bernal, A., \& Guti\'{e}rrez, L. 2014, RMxAA, 50, 323

\bibitem[Blondin et al.(1998)]{Blondin1998}Blondin, J. M., Wright, E. B., Borkowski, K. J., \& Reynolds, S. P. 1998, ApJ, 500, 342

\bibitem[Brantseg (2013)]{Brantseg2013}Brantseg, T.F. 2013, The University of Iowa,  Dissertations \& Theses

\bibitem[Caprioli et al. (2018)]{Caprioli2018}Caprioli, D.,  Zhang, H., \& Spitkovsky,A. 2018, Journal of Plasma Physics 84, 715840301

\bibitem[Caswell(1970)]{Caswell1970}Caswell, J. L. 1970, Austral. J. Phys., 23, 105

\bibitem[Clark \& Caswell (1976)]{Clark1976}Clark, D. H., \& Caswell, J. L. 1976, MNRAS, 174, 267

\bibitem[Cox (1972)]{Cox1972}Cox, D. P. 1972, ApJ, 178, 159

\bibitem[Cristofari \& Blasi  (2019)]{Cristofari2019}Cristofari, P., Blasi, P. 2019, MNRAS, 489, 108

\bibitem[Davies \& Meaburn(1978)]{Davies1978}Davies, R.D., Meaburn, J. 1978, A\&A, 69, 443

\bibitem[Day et al.(1972)]{Day1972}Day, G. A., Caswell, J. L., \& Cooke, D. J. 1972, Austral. J. Phys. Astrophys.
Suppl., 25, 1



\bibitem[Feng et al.(2019)]{Feng2019}Feng, L., Li, Z.-Y., Su, M., et al. 2019, RAA, 19, 046

\bibitem[Fesen et al.(1985)]{Fesen1985}Fesen, R. A., Blair, W. P., \& Kirshner, R. P. 1985, ApJ, 292, 29



\bibitem[Frail et al. (1996)]{Frail1996}Frail, D. A., Goss, W. M., Reynoso, E. M., et al. 1996, AJ, 111, 1651


\bibitem[Graham et al.(1982)]{Graham1982}Graham, D. A., Haslam, C. G. T., Salter, C. J., \& Wilson, W. E. 1982, A\&A, 109, 145 

\bibitem[Guo et al.(2017)]{Guo2017}Guo, X.-L., Xin, Y.-L., Liao, N.-H., et al. 2017, ApJ, 835, 42


\bibitem[H.E.S.S. Collaboration (2018)]{HESS2018}H.E.S.S. Collaboration, Abdalla, A., Abramowski, A., et al. 2018, A\&A, 612, A3

\bibitem[Holden(1968)]{Holden1968}Holden, D.J. 1968, MNRAS, 141, 57




 
\bibitem[Leahy(1986)]{Leahy1986}Leahy, D. A. 1986, A\&A, 156, 191

\bibitem[Morlino \& Caprioli(2012)]{Morlino2012}Morlino, G., \& Caprioli, D. 2012, in AIP Conf. Ser. 1505, 5th Int. Meeting on High Energy Gamma-Ray Astronomy (Melville, NY: AIP), 241

\bibitem[Nolan et al.(2012)]{Nolan2012}Nolan, P. L., Abdo, A. A., Ackermann, M., et al. 2012, ApJS, 199, 31

\bibitem[Nousek(1981)]{Nousek1981}Nousek, J. A., Cowie, L. L., Hu, E., Lindblad, C. J., \& Garmire, G. P. 1981, ApJ, 248, 152

\bibitem[Odegard(1986)]{Odegard1986}Odegard, N. 1986, ApJ, 301, 813

\bibitem[Reich et al. (1997)]{Reich1997}Reich, P., Reich, W., \& Furst, E. 1997, A\&AS, 126, 413

\bibitem[Su et al.(2017b)]{Su2017b}Su, Y., Zhou, X., Yang, J., et al. 2017b, ApJ, 836, 211

\bibitem[Tang et al.(2013)]{Tang2013}Tang, Y.-Y., Dai, Z.-C., \& Zhang, L. 2013, RAA, 13, 537


\bibitem[Xi et al.(2020a)]{Xi2020a}Xi, S.-Q., Liu, R.-Y, Wang, X.-Y., et al. 2020a, ApJL, 896, L33

\bibitem[Xi et al.(2020b)]{Xi2020b}Xi, S.-Q., Zhang, H.-M., Liu, R.-Y., \& Wang,X.-Y. 2020b, ApJ, 901, 158

\bibitem[Xing et al.(2016)]{Xing2016}Xing, Y., Wang, Z.-X., Zhang, X., \& Chen, Y. 2016, ApJ, 823, 44

\bibitem[Xiang \& Jiang (2021)]{Xiang2021a}Xiang, Y.-C., \& Jiang, Z.-J. 2021a, APJ, 908, 22

\bibitem[Yuan et al. (2018)]{Yuan2018}Yuan, Q., Liao, N.-H., Xin, Y.-L., et al. 2018, ApJL, 854, L18

\bibitem[Zabalza (2015)]{Zabalza2015}Zabalza, V. 2015, ICRC (The Hague), 34, 922

\bibitem[Zeng et al.(2017)]{Zeng2017}Zeng, H., Xin, Y., Liu, S., et al. 2017, ApJ, 834, 153

\bibitem[Zeng et al.(2019)]{Zeng2019}Zeng, H., Xin, Y., \& Liu, S. 2019, ApJ, 874, 50

\bibitem[Zhang \& Fang (2007)]{Zhang2007}Zhang, L., \& Fang, J. 2007, ApJ, 666, 247


\end{thebibliography}
\end{document}